\documentclass[pra,floatfix,10pt]{revtex4}
\usepackage{amsmath}
\usepackage[english]{babel}
\usepackage[pdftex]{graphicx}
\usepackage{color}
\usepackage{subfig}

\usepackage[sort&compress]{natbib}
\usepackage{hyperref}
\usepackage{amsmath}
\usepackage{amssymb}
\usepackage{mathtools}
\usepackage{braket}

\def\fcal{\mbox{$\cal F\,$}}

\begin{document}
\title{\bf  Photonic realization of the $\kappa$-deformed Dirac equation }

\author{P. Majari}
\email{majari@icf.unam.mx}
\affiliation{Instituto de Ciencias F\'isicas, Universidad Nacional Aut\'onoma de M\'exico, Cuernavaca 62210, M\'exico}

\author{E. Sadurn\'i}
\email{sadurni@ifuap.buap.mx}
\affiliation{Instituto de F\'isica, Benem\'erita Universidad Aut\'onoma de Puebla,
Apartado Postal J-48, 72570 Puebla, M\'exico}

\author{M. R. Setare}
\email{rezakord@ipm.ir}
\affiliation{Department of Science, Campus of Bijar, University of Kurdistan,
Bijar, Iran}

\author{J.  A.  Franco-Villafa\~{n}e}
\email{jofravil@ifisica.uaslp.mx}
\affiliation{CONACYT-Instituto  de  F\'isica,  Universidad  Aut\'{o}noma  de  San  Luis  Potos\'i,  78290  San  Luis  Potos\'i, SLP.,  M\'exico}

\author{T. H. Seligman}
\email{seligman@icf.unam.mx }
\affiliation{Instituto de Ciencias F\'isicas, Universidad Nacional Aut\'{o}noma de M\'exico, Cuernavaca 62210, M\'exico,\\
 Centro Internacional de Ciencias, Cuernavaca-62210, M\'exico}

\date{\today}% It is always \today, today,

\begin{abstract}
\noindent
We show an implementation of a $\kappa$-deformed Dirac equation in tight-binding arrays of photonic waveguides. This is done with a special configuration of couplings extending to second nearest neighbors. Geometric manipulations can control these evanescent couplings. A careful study of wave packet propagation is presented, including the effects of deformation parameters on Zitterbewegung or trembling motion. In this way, we demonstrate how to recreate the effects of a flat noncommutative spacetime --i.e., $\kappa$ Minkowski spacetime-- in simple experimental setups. We touch upon elastic realizations in the section of Conclusions.
\hspace{0.35cm}

\end{abstract}

\maketitle
\section{Introduction}

The field of quantum simulations has been of pivotal importance in the study of physical systems that are experimentally out of reach or beyond natural observation. Notable examples in cold matter can be found in \cite{R16, R17} and \cite{R27, R29, R31} using trapped ions. In this paper, we address the possibility of emulating the effects of Lorentz algebraic deformations on the motion of relativistic electrons \cite{R40, R41}. In theory, such deformations are associated with a noncommutative geometry of spacetime and a generalized uncertainty principle \cite{R37, RXX, R30, R32}. On physical grounds, the hypothetical corrections stem from  a fundamental length scale, which is of a quantum mechanical nature.

Let us describe the general features of our emulations to put our work into context. Within the class of complex quantum simulations, there are simpler systems whose properties can be studied with single body dynamics. These systems, in turn, contain all mesoscopic realizations of quantum mechanical wave equations in microwaves \cite{R8, R14, R18, R24, p25, p26} and photonics \cite{R5, R6, R11, R12, R15, R35, R68, R33}. For optical fiber couplings see \cite{R1, R7, R4, R34}. Such table-top experiments provide flexible configurations and easy tuning of parameters, including recent constructions of effective relativistic systems. Examples of condensed matter realizations and their emulations also abound \cite{R57, R66, RYY, R19, R20, R21, R22}.
The success of dynamical analogies between quantum mechanical equations and electromagnetic waves in various important subjects --quantum graphs, chaotic scattering and billiards, tight-binding arrays and crystals, including graphene and phosphorene-- has led us to consider new experimental challenges towards high-energy extensions of the Dirac equation, ruling the motion of ultra relativistic fermions. Indeed, Dirac Hamiltonians in 1+1 and 2+1 dimensions have been produced effectively in a variety of tight-binding systems supported by honeycomb lattices, and in more general settings, by any bipartite (spinorial) lattice entailing conical energy relations, also dubbed Dirac points.

These precedents become important when we look at the so-called $\kappa$-deformed algebras at hand \cite{R40, R41, R43, R49} and even in the field of $q$ deformations \cite{R39, R42}. Their effects on quantum field theory have been carefully studied \cite{R50, R51, R52, R53, R54, R55, R62, R38} including particle statistics \cite{R64}. In this regard, it has been proved that quadratic corrections in the momentum of a particle will modify the usual Dirac Hamiltonian defined in empty space, even when it represents a physical situation free of interactions \cite{R60, R61, R65}. As previously mentioned, this is known to take place in the presence of a postulated minimal length, presumably in the order of $G \hbar / c^3 \sim 1.6 \times 10^{-33}$ cm, i.e., the Planck scale. The corresponding corrections would imply a modified energy-momentum relation and extraordinary dispersion relations of matter waves propagating under the effects of new physics. The idea is to emulate them.

In connection with fundamental aspects of physics, some words are in order. Presumably, the existence of a fundamental length is the outcome of a foamy space consistent with theories that deal with the very nature of spacetime or its emergent properties starting from string theory. However, it is still unknown whether such structures actually underlie our physical world, and it is even more uncertain whether we shall be able to observe the actual consequences of their existence in high energy experiments or cosmological observations. Important efforts in the phenomenology of deformations and minimal lengths can be found in \cite{R44, R45, R47}, including a plausible stringy origin \cite{R46} and Planck scale phenomenology in \cite{R36, R48}.

For this reason, we recreate here the conditions in which the aforementioned effects can be observed. Our aim is to engineer the corresponding dispersion relations with a tight-binding scheme consistent with previous successful emulations of relativistic wave equations. In what touches wave propagation, our emulations shall be able
to produce two important outcomes effectively: i) A modified energy spectrum in accordance with predictions from potentially new physics and ii) a corrected evolution of wave packets with modified group velocities in empty space. The first result can be easily achieved by introducing second-neighbor interactions --or hopping amplitudes-- in a crystal where Dirac points are initially ensured; this shall be done by simple geometric manipulations of resonators in various realizations, such as optical fibers and ceramic disks. The second result will be tested by a close inspection of a phenomenon known as Zitterbewegung \cite{R26}, already produced artificially in Dirac lattices \cite{R13, R16, R17, R29} and calculated in previous treatments \cite{R58, R59}, where some of them cover carefully the full energy band \cite{R28}. This reaches well beyond the conical region of the emulated spectrum. With our treatment, we shall be able to compare the oscillation frequency of a wavepacket's width -- as well as its decay in amplitude -- with the expected theoretical predictions, finding thereby new significant  effects coming from a hypothetical minimal length, together with corrected trajectories of electrons obtained as average positions in the $\kappa$-deformed Heisenberg picture.\\
\\
Structure of the paper: In section \ref{sec:2} we revisit the emergence of the $\kappa$ deformed Dirac equation and obtain thereby the first corrections in the Dirac Hamiltonian due to a fundamental length $a$. In section \ref{sec:3} we present a careful construction of arrays made of coupled optical fibers disposed in a strip resembling a triangular lattice, fulfilling thus a Dirac-like dynamical equation with conical points. In section \ref{sec:4} we study the effects of the deformation on the trembling motion of wave packets, including the corrections in the width coming from $a$. A detailed full-band computation of Zitterbewegung for tight-binding arrays with second neighbors is offered in \ref{sec:4.1}. We conclude in \ref{sec:5}.

 \section{ The $ \kappa$-deformed Dirac equation \label{sec:2}}

Noncommutative (NC) geometry was envisaged by Gelfand when he showed that a space is determined by the algebra of the functions acting on it. The notion of space is then tied to the nature of its algebra; therefore a noncommutative spacetime (NCST) follows from a noncommutative algebra. This has been extensively studied in the framework of Hopf algebras and the so-called quantum groups \cite{R49, R50, R51}. In our case, the coordinate functions  $x_\mu$,  satisfy the commutation relations of the form $[x_\mu,x_\nu]=i (\Theta_{\mu\nu}+\Theta_{\mu\nu}^\lambda x_\lambda+...)=i \Theta _{\mu\nu}(x)$, a relation that has been instrumental in the construction of a deformed quantum field theory \cite{R52}. Among all the possible ways of representing NCST, we are particularly interested in the $\kappa$-Minkowski spacetime employed in \cite{R53,R54,R55,R52}. In other words, the $\kappa$-Minkowski spacetime is a Lie algebraic deformation of the usual Minkowski (flat) spacetime where the deformation parameter can be related to a length scale in which quantum gravity might take place. The corresponding $\kappa$-Poincar\'e-Hopf algebraic relations can  be written in terms of a deformation parameter $\kappa=1/|a|$,  as $[\hat{x}_\mu,\hat{x}_\nu]=i(a_\mu \hat{x}_\nu-a_\nu \hat{x}_\mu)$. Moreover, the existence of such a fundamental scale can be encoded in Dirac operators acting on spinor fields. The deformed Dirac equation obtained in the  framework of the $\kappa$-Poincar\'e-Hopf algebra and its equivalent in periodic arrays of coupled waveguides will be studied in the following.

Let us start with the algebra related to NC spaces satisfying the following relations \cite{R60, R61}:

\begin{equation}\label{1}
[M_{i0},\hat{x}_0]=-\hat{x}_i+ia M_{i0},
\end{equation}

\begin{equation}\label{2}
[M_{i0},\hat{x}_j]=-\delta_{ij}\hat{x}_0+ia M_{ij},
\end{equation}
where $M_{\mu\nu}$  contains the rotation and boost generators  of the $\kappa$-Poincar\'e algebra and $\hat{x}_\mu$ denotes the NC coordinates. This algebraic structure also entails the following relations:
\begin{equation}\label{3}
[\hat{x}_\mu,\hat{x}_\nu]=iC_{\mu \nu \lambda}\hat{x}^\lambda=i(a_\mu \hat{x}_\nu-a_\nu \hat{x}_\mu),
\end{equation}
where the structure constants are written in terms of a Minkowski vector $a_\mu$ and the flat metric $\eta_{\mu \nu}$: $C_{\mu\nu\lambda}=a_\mu \eta_{\nu\lambda}-a_\nu \eta_{\mu\lambda}$. In some frame of reference  $a_i=0$, $a_0=a$ and $\hat{x}_i=x_i \phi$ where $\phi$ is so far free. In the $\kappa$-Poincar\'e algebra, the modified derivative operators $D_\mu$, the so-called
Dirac derivatives, are given by \cite{R62, R51}:
\begin{equation}\label{4}
D_0=\partial_0 \left({\sinh(A)\over A}\right)+{i a \nabla^2 e^{-A}\over 2 \phi^2},
\end{equation}
and
\begin{equation}\label{5}
D_i=\partial_i{ \left({e^{-A}\over \phi}\right)},
\end{equation}
with $A=-ia \partial_0$ and $a=\kappa^{-1}$. This leads to the following relations \cite{R64}:

\begin{equation}\label{6}
[M_{\mu\nu},D_\lambda]=\eta_{\nu \lambda}D_\mu-\eta_{\mu \lambda}D_\nu,
\end{equation}

\begin{equation}\label{7}
[D_\mu,D_\nu]=0,
\end{equation}
where the metric's signature convention is fixed as $\eta_{\mu\nu}= \rm{diag}(-1,1,1,1)$. In such a way, the deformed Dirac equation is now postulated in terms of $D_{\mu}$ as

\begin{equation}\label{9}
\left[i \gamma^\mu D_\mu+m \right]\psi=0.
\end{equation}
With the special choice  $\phi=e^{-A}$,  we eliminate the deformation in the spatial derivatives, leaving us only with a new (corrected) time component $D_0$. We note here that this choice, together with the definition of $A$, turns $\phi$ into a nonlocal operator in time. By substituting (\ref{4}), (\ref{5})  in the above equation  and after a few straightforward manipulations,  the following $\kappa$-Dirac  equation is deduced:

\begin{equation} \label{9.1}
\left[i \gamma^0 \left(\frac{i}{a} \sinh(A) +{ia\over 2}\nabla^2 \right)+i \gamma^i \partial_i+m \right]\psi=0
\end{equation}
and this equation is, in fact, nonlocal due to the obvious relation $\phi \psi(t) = \psi(t + i a) $ for any wave function $\psi$. On physical grounds, we may take only the first corrections in $a$ with the aim of describing a slightly perturbed Dirac operator (note however, that this concession is not made on mathematical grounds, because infinite order differential equations cannot be truncated without dire consequences on the oscillatory behavior of their solutions) leading to

\begin{equation}\label{10}
\left[i \gamma^0 \left(\partial_0+{ia\over 2}\nabla^2 \right)+i \gamma^i \partial_i+m \right]\psi=0.
\end{equation}
It follows from the above  equation that the corresponding Hamiltonian is

\begin{equation}\label{11}
H = c \alpha \cdot p +{ a \over {2 }}\nabla^2 + m \beta.
\end{equation}
Here, the explicit representation of Dirac matrices in $1+1$ dimensions given by $\alpha_1 = \sigma_1$ and $\beta=-\sigma_3$ is possible and it is consistent with our choice of metric signature.
It is also obvious that the undeformed Dirac equation is obtained in the limit $a\rightarrow 0$.  However, we underscore the fact that, experimentally, the deformation parameter $a$ is greatly limited by an upper bound of the order of $a <10^{-29}$ m \cite{R66}.

Several effects can be investigated using this new Hamiltonian. It might well be that the presence of $a$ modifies the spectrum of a relativistic particle, but it is not easy to gain access to such energy scales in accelerators. There are other effects that could be amplified in other settings, e.g. wave packet evolution. We shall explore this possibility in further sections.

\section{An array of photonic waveguides \label{sec:3}}

The tight-binding model of the $\kappa$-deformed Weyl equation can be implemented on a macroscopic experiment using an array of microwave resonators, as shown in figure \ref{sh2}. The resonators can be built as cylinders of the same size, but with two different dielectric constants, for example, Exxilia Temex Ceramics E2000 and E3000 with $\epsilon=36$ and $34$, respectively. An induced mass parameter $\alpha$ around $0.28$~GHz is expected for cylinders of $8$~mm diameter, and their length much larger than their diameter. The coupling parameter $\xi$ between the cylinders can be set between $0.4$~GHz and $1$~GHz, depending on the separation between cylinders. With those experimental parameters, it is possible to estimate the Zitterbewegung characteristic frequencies $\omega'$ and $\omega$. The corresponding Zitterbewegung oscillation lengths that we named here $\lambda'$ and $\lambda$ respectively will be of order $\lambda'\sim$~cm and $\lambda\sim$~m. Those scales make the Zitterbewegung effect observable on a macroscopic scale, thanks to our tight-binding representation of the corresponding wave operator. In order to excite the resonators, we propose to use an array of antennas. Each antenna will be placed near the end of each fiber or cylinder. The antennas should all be parallel and oriented in such a way that they are capable of exciting an electric field perpendicular to the optical axis and to the horizontal axis shown in fig. \ref{sh2}.
The array of antennas has to be excited by the same microwave frequency but, at the same time, allowing the control of the input power independently in each antenna. One possible and inexpensive way to feed the antennas is to use a direct digital synthesizer (DDS) that provides independent frequency, phase, and amplitude control on each channel. The propagation of an optical field  of  disordered waveguide arrays by using tight-binding approximation is given by:

\begin{figure}
 \centering
 {\includegraphics[width=0.50\textwidth]{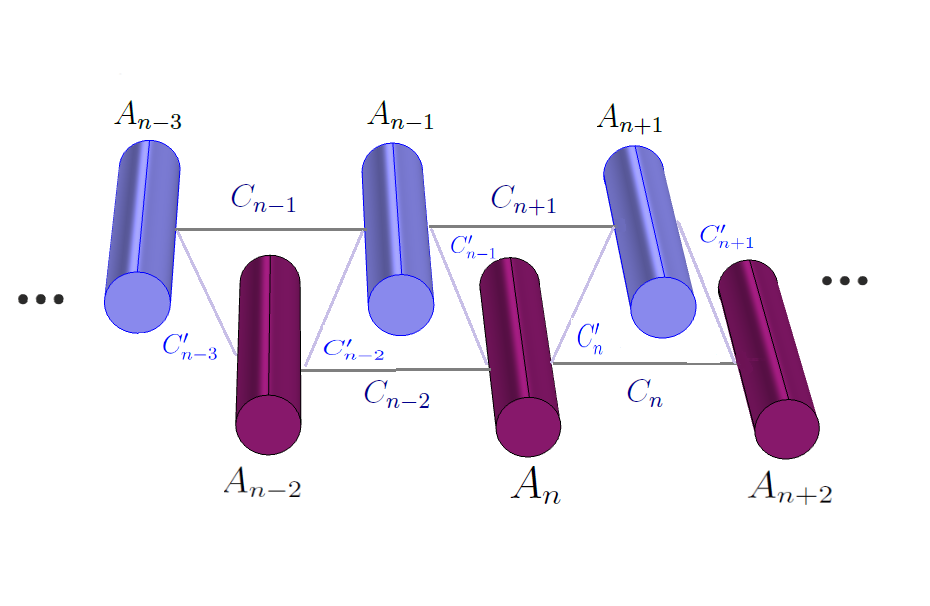}}
  \caption{Schematic view of a binary array made of two types of waveguides, A and B, arranged in a triangular lattice along a strip.
 }\label{sh2}
\end{figure}

\begin{equation}\label{12}
i{ d E_n\over dz}+{(-1)^n}\alpha E_n+C_{n-2}E_{n-2}+C_{n}E_{n+2}+C^\prime_{n}E_{n+1}+C^\prime_{n-1}E_{n-1}=0,
\end{equation}
where $E_n\equiv E(n,z)$ is the electric field amplitude at the nth waveguide. Next, we let upper waveguide array  have odd numbers. Then

\begin{equation}\label{p14}
i{ d E_{2n}\over dz}+\alpha E_{2n}+C_{2n-2}E_{2n-2}+C_{2n}E_{2n+2}+C^\prime_{2n}E_{2n+1}+C^\prime_{2n-1}E_{2n-1}=0,
\end{equation}
and let the lower array label by even ones:

\begin{equation}\label{p15}
i{ d E_{2n-1}\over dz}-\alpha E_{2n-1}+C_{2n-3}E_{2n-3}+C_{2n-1}E_{2n+1}+C^\prime_{2n-1}E_{2n}+C^\prime_{2n-2}E_{2n-2}=0.
\end{equation}
Now by setting $E_{2n}=(-1)^n \psi_1(n,z)\equiv (-1)^n \psi_1(n)$ and $E_{2n-1}=-i(-1)^n \psi_2(n,z)\equiv -i(-1)^n \psi_2(n)$, Eqs   (\ref{p14})  and (\ref{p15}) can  be written as:

\begin{equation}\label{p16}
i{ d \psi_1(n)\over dz}+\alpha \psi_1(n)-C_{2n-2}\psi_{1}(n-1)-C_{2n}\psi_{1}(n+1)+iC^\prime_{2n}\psi_{2}(n+1)-iC^\prime_{2n-1}\psi_{2}(n)=0,
\end{equation}
and

\begin{equation}\label{p17}
i{ d \psi_2(n)\over dz}-\alpha \psi_2(n)-C_{2n-3}\psi_{2}(n-1)-C_{2n-1}\psi_{2}(n+1)+iC^\prime_{2n-1}\psi_{1}(n)-iC^\prime_{2n-2}\psi_{1}(n-1)=0,
\end{equation}
It is straightforward to show that  by considering $ C_{2n\pm i}=\eta$ and $C^\prime_{2n-i}=\xi$ with $i=0,1,...$ these equations  reduce to

\begin{equation}\label{0}
i{ d \over dz}\begin{bmatrix}
\psi_1(n)\\
\psi_2(n)
\end{bmatrix}
=\begin{bmatrix}
-\alpha \psi_1(n)+\eta\psi_1(n-1)+\eta \psi_1(n+1)-i \xi \psi_2(n+1)+i \xi \psi_2(n) \\
+\alpha\psi_2(n)+\eta \psi_2(n-1)+\eta \psi_2(n+1)-i \xi \psi_1(n)+i \xi \psi_1(n-1)
\end{bmatrix}
=H\begin{bmatrix}
\psi_1(n)\\
\psi_2(n)
\end{bmatrix}
\end{equation}

with a Hamiltonian operator defined as

\begin{equation}\label{ab0}
{{H}}=\begin{bmatrix}
-\alpha +\eta T^{-1}+\eta T& -i\xi T+i\xi\\
-i\xi+i\xi T^{-1}  & \alpha +\eta T^{-1}+\eta T
\end{bmatrix},
\end{equation}
where $T$  is the translation operator in one unit of $n$. In order to reduce the system of equations to a $2 \times 2$ matrix, we must note that in case of Bloch wave transport inside the fibers, the wave function must be written in the form $(\psi_1(n,z),    \psi_2(n,z))^\dag= (A(z)e^{ikn},    B(z)^{ikn})^\dag$ which  satisfies Bloch's
theorem $T e^{ikn}=e^{ik(n+1)}$. Now, the Hamiltonian \ref{ab0} is reduced to

\begin{equation}\label{p18}
{{H}}=\begin{bmatrix}
-\alpha +2 \eta \cos(k)& -i\xi e^{ik} +i\xi\\
-i\xi+i\xi e^{-ik}  & \alpha +2 \eta \cos(k)
\end{bmatrix}
\simeq \begin{bmatrix}
-\alpha+2\eta-\eta k ^ 2& \xi {k} \\
\xi {k}   & \alpha+2\eta-\eta k ^ 2
\end{bmatrix}
\end{equation}
 Therefore, the energy eigenvalues are given by the formula

\begin{equation}\label{p19}
E=2\eta \cos(k)+s \sqrt{4\xi ^ 2 \sin^ 2 \left({k\over 2} \right)+\alpha^ 2}
\end{equation}
where $s=\pm 1$. This holds even beyond the Dirac (conical) points. The eigenfunctions are two-component spinors of the form

\begin{equation}\label{p19}
\binom{u_1}{u_2}
= {1\over \sqrt{2(E-2\eta \cos(k))}}\binom{{\sqrt{E-\alpha-2\eta \cos(k)} }}{{e^{-ik/2}\sqrt{E+\alpha-2\eta \cos(k)} }}.
\end{equation}

By setting $k\rightarrow p_x$, i.e. in units where $\hbar =1$, we obtain:

\begin{equation}\label{p20}
H= {-\eta {{p_x}^ 2}}I_2+ \xi p_x \sigma_x-\alpha \sigma_z +2\eta I_2=H_0+V
\end{equation}
where $V= 2\eta$ is a constant potential, therefore irrelevant in the dynamics. Finally, we note that after the formal change ${  a \over 2 }\rightarrow {\eta}$,  $m \rightarrow \alpha$  and $1 \rightarrow \xi$,  the expression for the Hamiltonian in (\ref{11}) can be mapped to $H_0$ previously written in (\ref{p20}).

\section{Evolution of position in $\kappa$-deformed Dirac theory \label{sec:4}}

Now we would like to investigate one of the special features of the Dirac equation, the trembling motion known as Zitterbewegung. To clarify this effect in the $\kappa$-Dirac equation, we must calculate the time  evolution of the position operator under the strict conditions $\eta \neq 0$, $\alpha \neq 0$. In the absence of rest mass (Weyl equation) we know that there is no visible effect, for the evolution of $x$ would be trivial. The calculation of $x(t)$ for the more general case $\alpha \neq 0$ is, however, straightforward and we shall proceed in this direction. In the Heisenberg picture, we have \cite{R66}:

\begin{equation}\label{17}
x(t)=x(0)-\eta p_x t+\xi^2 p_x {(H^\prime)}^{-1}t+{i  \xi \over 2 } {H^\prime}^{-1}\left[\sigma_x {-} \xi{p}_x( {H^\prime})^{-1} \right](e^{{-2i {H^\prime}t }}-1),
\end{equation}
where ${H^\prime}= -\alpha \sigma_z + \xi \sigma_x {p_x}  $.
To see the dependence on the deformation parameter more clearly, we focus now on the width of wave packets $\langle (\Delta x)^2 \rangle = \langle x \rangle^2 - \langle x^2 \rangle $. Due to Ehrenfest's theorem, the time average $\langle x \rangle$ suffers the same modifications as the classical trajectory of a particle governed by a $\kappa$-deformed energy momentum relation. On the other hand, the second term $\langle {x^2}\rangle_\psi$ provides an important modification to the wave-like behavior of the particle. Its explicit form is given by

\begin{equation}\label{p24}
\langle x^2 \rangle_\psi= \fcal t^2 -\sqrt{2\pi t\over \tau}\left({2 \eta \xi^2 \alpha\pi^2 \sin({\omega ^\prime t } ) \over {\omega ^\prime}^3}\right)\Gamma+
{1\over \sqrt{t}}\left[\sqrt{2\pi\over \tau}\left({ \xi^2 \alpha^2 \sin({\omega ^\prime t } )^2 \over {\omega ^\prime}^4}\right)\Lambda_{1}
+\sqrt{2\pi\alpha\over \xi^2}\left({ \xi^2 \alpha^2 \sin({\omega  t } )^2 \over {\omega}^4}\right)\Lambda_{2}\right]\hspace{10.1em}
\end{equation}
as shall be derived later on.

\begin{figure}
\centering
\subfloat[]{\includegraphics[width=0.50\textwidth]{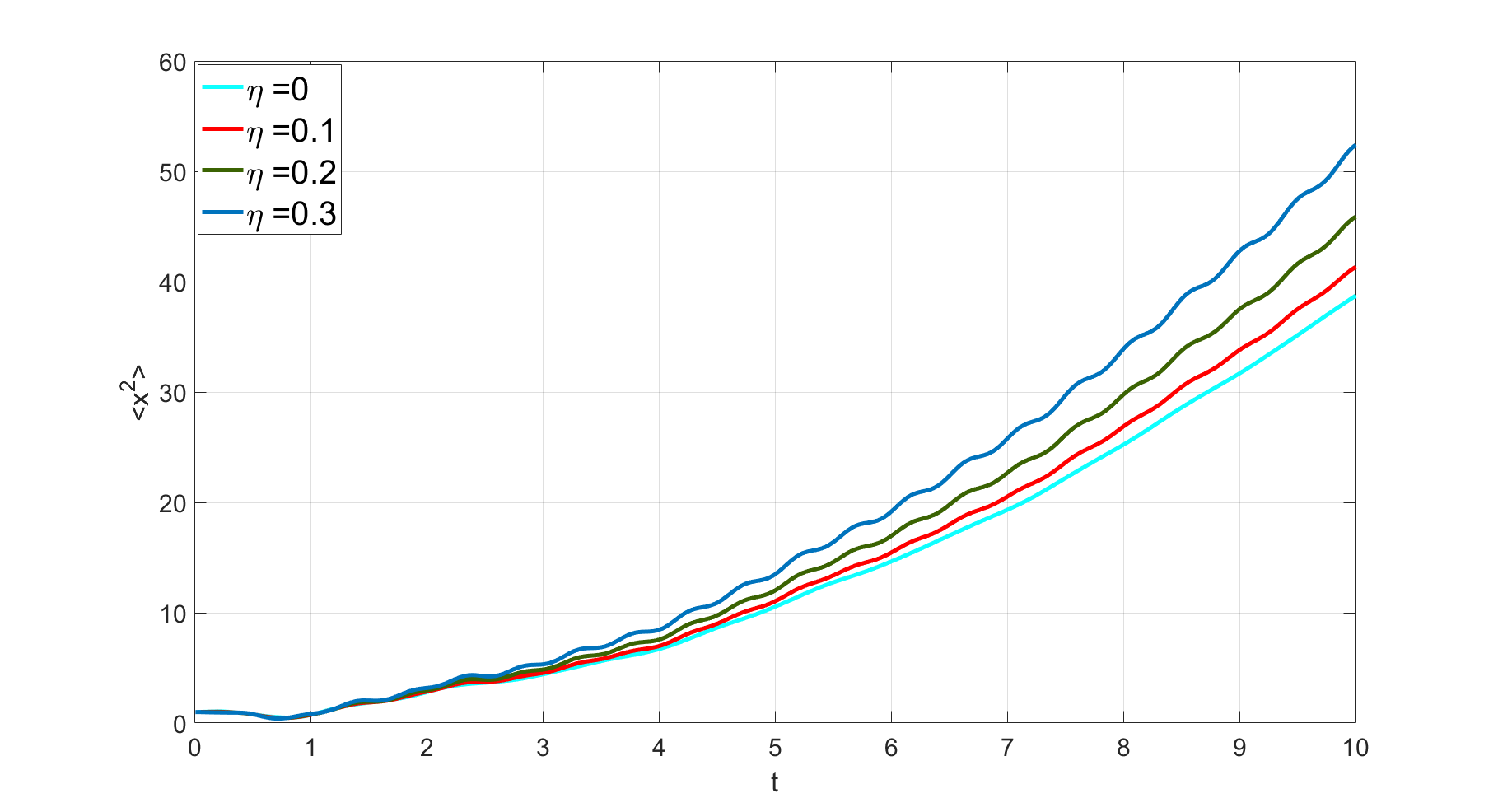}}
  \hfill
  \subfloat[]{\includegraphics[width=0.50\textwidth]{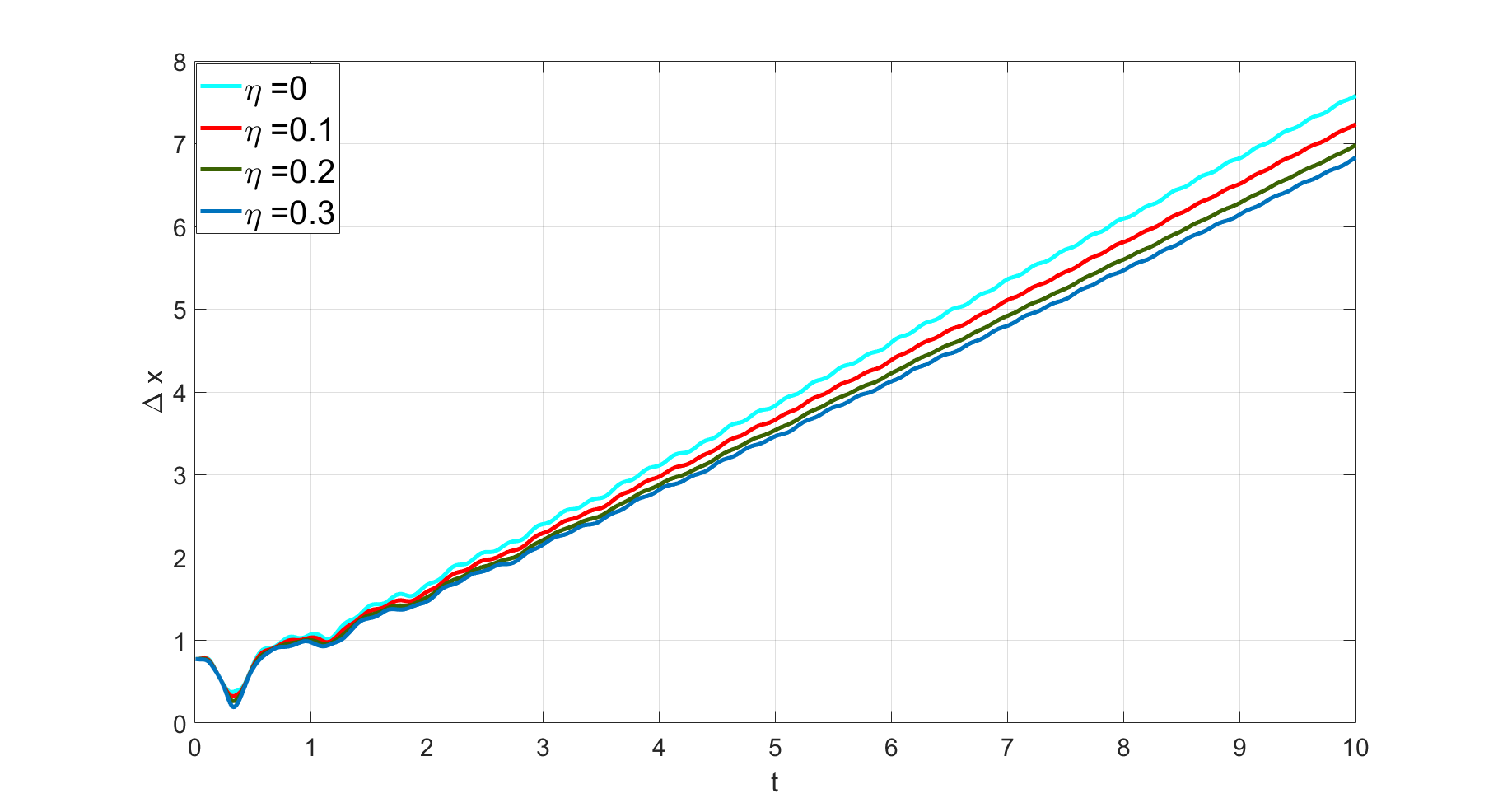}}
  \caption{Time development of (a) $\langle{x^2}\rangle_\psi$   and (b) variance with  $ \xi=\alpha=4 $  and  a few values of $\eta$. }\label{sh2}
\end{figure}

The time development of $ \langle{x^2}\rangle_\psi$, with $x(0)=1.5$ and initial width 0.8 is shown in fig. (\ref{sh2}).
The interesting effect that has been obtained consists of a deformation-dependent evolution of a particular component in the width contributing to the usual ballistic expansion. In general, we can appreciate in fig.\ref{sh2}(b)  that wavepackets spread faster due to the deformation, as shown by $\fcal$. We also see in the results shown in fig. (\ref{sh2}) (a) that the damping of the trembling part disappears with the envelope $1/\sqrt{t}$, but some of its oscillations are 'prolonged' by an increase of $\eta$, from 0 to 0.3. Indeed, the envelope $\sqrt{t}$ is affected by $\eta$ in the second term of (\ref{p24}), while the rest of the contributions remain unperturbed due to a phase cancellation of the $\eta$ component of the energy. This direct proportionality in $\Gamma$ amplifies the phenomenon in time, but in case of realistic values of $\eta$ within experimental bounds, it would be too challenging to detect them in experiments with electrons. In optical realizations, such a parameter is at our disposal, with recommended values shown in the inset of fig. \ref{dr2} and \ref{sh2} (b), for a better appreciation.

\subsection{ Zitterbewegung in the photonic lattice: computations \label{sec:4.1}}
\begin{figure}
{\includegraphics[width=0.50\textwidth]{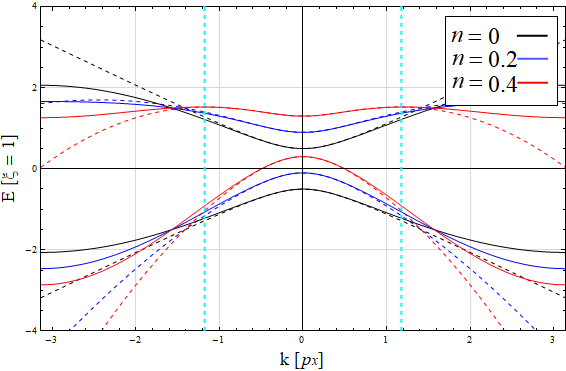}}
  \caption{A comparison of dispersion relations for waves governed by our tight-binding array (solid lines) and the $\kappa$-deformed Dirac equation (dashed lines). Mass $\alpha=0.5$, coupling $\xi = 1$. There is good agreement near the Dirac point at $k=0$. Deformation parameters: $\eta =$ 0 (black) 0.2 (blue) 0.4 (red). The tight-binding relation displays new points of vanishing group velocity in the upper band, marked by light blue vertical lines. The phase factors, however, cancel out in the computation of averages. }\label{dr2}
\end{figure}

We derive the time evolution of position operator for photonic waveguide arrays. It is important to do so without approximations in the tight-binding dispersion relations for an honest comparison with deformed theories. Since the stationary phase approximation will be required in the derivation of averages, it is important to analyze the energy landscape in (Bloch) momentum space or Brillouin zone in search for vanishing group velocities. A comparison of energy curves for some values of $\eta$ is given in fig. \ref{dr2}.

The time average of the position using the Dirac theory (\ref{17}) can be written as

\begin{equation}\label{p21}
\langle x_{ZB}\rangle_\psi=\left\langle \left\{ {i  \xi \over 2 } {H^\prime}^{-1}[\sigma_x {-} \xi{p}_x( {H^\prime})^{-1} ](e^{{-2i {H^\prime}t }}-1) \right\} \right\rangle_\psi,
\end{equation}
but using the wave packet decomposition with Fourier coefficients $\psi_{k,s}$, one also has

\begin{eqnarray}\label{p22}
\langle x_{ZB}\rangle_\psi &=&  \alpha^2 \xi^2\sum _{s} \int_{0}^{\pi} dk {k  \sin({2E ^\prime t})\over 2 {E ^\prime}^4 } |\psi_{k,s}|^2+
{  \xi }\sum _{s,s^\prime} \int_{0}^{\pi} dk \left[  {i \xi\alpha\over {E ^\prime}^2 }\sin^2({E ^\prime t})({u_1}^*u_2-{u_2}^*u_1)- \right. \\ &-&
\left. { \xi\alpha^2\over 2 {E ^\prime}^3 }\sin({2E ^\prime t})({u_1}^*u_2+{u_2}^*u_1)  \right](\psi_{k,s}{\psi_{k,s^\prime}}^*)\hspace{10.1em}
\end{eqnarray}
where $E^\prime=s \sqrt{4\xi ^ 2 \sin^ 2({k\over 2})+\alpha^ 2}$.  Now  by using
the stationary phase approximation for $k=0,\pi$ we obtain

\begin{equation}\label{p23}
\langle x_{ZB}\rangle_\psi\simeq \left({2  \pi \over t|{{\xi}^2\over \omega ^\prime }| }\right)^{1\over 2}e^{{-it \omega ^\prime}}F(\alpha,\xi)+
\left({2  \pi\over t|{\xi ^2\over \omega}| }\right)^{1\over 2}e^{{-it \omega}}G(\alpha,\xi)+
\left({  \pi\over t|{\xi ^2\over \omega^\prime}| }\right)^{1\over 2}e^{{-it \omega^\prime\over 2}}L(\alpha,\xi)
+ \rm{c.c.}
\end{equation}
where $\omega ^\prime=s\sqrt{4\xi ^ 2 +\alpha^ 2}$ and $\omega = s\alpha$. It is clear that the trembling motion vanishes with an envelope curve $1/\sqrt{ t} $  \cite{R28}. We note that the amplitudes  and frequencies of oscillation are independent of
the deformation parameter. This result is  valid only for the Dirac approximation and, in general, the corrections due to $\eta$ have an impact on $\langle x_{ZB}\rangle$ for the tight-binding system with second neighbors. On the other hand, the width of the wave packet needs the average of the squared position operator $\langle{x^2}\rangle_\psi$ which is given by

\begin{equation}\label{p24}
\langle x^2 \rangle_\psi= \fcal t^2 -\sqrt{2\pi t\over \tau}\left({2 \eta \xi^2 \alpha\pi^2 \sin({\omega ^\prime t } ) \over {\omega ^\prime}^3}\right)\Gamma+
{1\over \sqrt{t}}\left[\sqrt{2\pi\over \tau}\left({ \xi^2 \alpha^2 \sin({\omega ^\prime t } )^2 \over {\omega ^\prime}^4}\right)\Lambda_{1}
+\sqrt{2\pi\alpha\over \xi^2}\left({ \xi^2 \alpha^2 \sin({\omega  t } )^2 \over {\omega}^4}\right)\Lambda_{2}\right]\hspace{10.1em}
\end{equation}
where the following shorthands are used:

We introduce the variable $\tau={{\xi^2  \over \sqrt{\alpha^2+\xi^2 \pi^2 }}-{\xi^4 \pi^2 \over {(\alpha^2+\xi^2 \pi^2)^{3/2}}}}$. The amplitude proportional to $t^2$ is given by
\begin{eqnarray}\label{p25}
\fcal &=&\sum_s {\int_0}^\pi dk  \left({k^2 \alpha^2 \xi^4 \over {E^\prime}^4}+{k^4  \xi^6 \over {E^\prime}^4}+k^2 \eta^2\right)|\psi_{k,s}|^2+
 \sum_s {\int_0}^\pi dk \left({2k^2 \eta \alpha \xi^2 \over {E^\prime}^2}\right)({u_1}^*u_1-{u_2}^*u_2)|\psi_{k,s}|^2-\\
 &-&\sum_{s,s^\prime} {\int_0}^\pi dk \left({2k^3  \eta \xi^3\over {E^\prime}^2}\right)({u_1}^*u_2+{u_2}^*u_1)\psi_{k,s}\psi_{k,s^\prime},\hspace{29.1em}
\end{eqnarray}
while the coefficient $\Gamma$ of the $\sqrt{t}$ envelope is
\begin{equation}\label{p26}
\Gamma=e^{-\pi/10}\left[\sin({\omega ^\prime t}+{\pi \tau\over 4})+{\alpha \over{\omega ^\prime} }\cos({\omega ^\prime t}+{\pi \tau\over 4})\right].
\end{equation}
Finally, the coefficients $\Lambda_{1,2}$ in the brackets of (\ref{p24}) (ruled by the $1/\sqrt{t}$ envelope) are as follows:

\begin{equation}\label{p27}
\Lambda_{1}=e^{-\pi/10}
\left[ \sin({\omega ^\prime t})\sin({\omega ^\prime t}+{\pi \tau\over 4})+ \frac{\cos(\omega ^\prime t+{\pi \tau\over 4})   \cos({\omega ^\prime t}) } {(\omega^\prime)^2}   (\alpha^2+\xi^2 \pi^2) \right]
\end{equation}
and
\begin{equation}\label{p28}
\Lambda_{2}=\sin({\omega  t}) \sin \left({\omega  t}+{\pi \xi^2\over 4\alpha}\right)+\alpha^2{  {\cos({\omega t}+{\pi \xi^2\over 4\alpha})   \cos({\omega  t})\over {\omega^2}}}.
\end{equation}

These results support our previous discussion on wave packet expansion.

\section{Conclusions \label{sec:5}}

We have succeeded in our emulation and theoretical study of deformed Dirac equations by means of photonic waveguide arrays. Once more, our tight-binding approach to coupling engineering has led to satisfactory results regarding spectrum and wavefunction simulation. We have shown how to surround subtle obstacles regarding the correspondence of photonic Zitterbewegung and relativistic trembling motion, as they differ by small but visible amounts when full frequency-band computations are employed. We were able to confirm the $\eta$ corrections, due to algebraic deformations, in the evolution of localized wave packets. Interestingly, at the end of the day, it was the $\sqrt{t}$ envelope what bore the $\kappa$ deformation, leading to a persistent oscillatory effect in the width and prolonged by an increase of $\eta$ as a direct proportionality. All other trembling components of the width remained untouched, including the rarely seen --i.e. short-lived but always present-- envelope $1/\sqrt{t}$. We also found a strong dependence on $\eta$ of the ballistic part $t^2$, controlling the overall speed of expansion, but such an effect already appears in the evolution of scalar particles, as it has little to do with spin.

From a technological point of view, we have shown that photonic waveguides may enable experimentalists to study
the effects of noncommutative spacetime in the lab. We should also comment on a renewed interest in elastic systems due to the flexibility of their experimental setups. The construction of elastic waveguides using Aluminum plates makes acoustic transport an attractive area in which emulations may play an interesting role, given the rich phenomenology of vibrational transport using various types of polarizations \cite{R2, R3, ZZ, R10}. As we mention above, microwave experience may also be consider for the realization of  $\kappa$-deformed Dirac equation.

\section*{Acknowledgments}
PM ~gratefully acknowledges a fellowship from UNAM-DGAPA. JAFV acknowledge   financial   support  from CONACYT project  A1-S-18696. THS and PM   acknowledge financial   support    from CONACYT project Fronteras 952, CONACYT project 254515 as well as UNAM-PAPIIT projects AG100891 and  IN113620.

\end{document}